\documentclass[12pt]{article}
\usepackage{amsfonts,amssymb}

\newcommand{\C}[1]{\mathbb{C}^{#1}}

\newcommand{\R}[1]{\mathbb{R}^{#1}}

\newcommand{\beq}{\begin{eqnarray}}
\newcommand{\eeq}{\end{eqnarray}}

\def \bi{\bibitem}
\def\){\right)}
\def\({\left( }

\newcommand{\vev}[1]{\langle#1\rangle}
\def\be{\begin{equation}}
\def\ee{\end{equation}}
\def\bea{\begin{eqnarray}}
\def\eea{\end{eqnarray}}
\newcommand{\eref}[1]{(\ref{#1})}

\newcommand{\rem}[1]{}

\def\ZZ{{\mathbb Z}}
\def\S{{\bf S}}

\def\Tr{{\rm Tr}}

\def \bi{\bibitem}
\def\np {  {\em Nucl. Phys.} }

\def \pr  { {\em Phys. Rev.} }

\def\ltap{\ \raise.3ex\hbox{$<$\kern-.75em\lower1ex\hbox{$\sim$}}\ }
\def\gtap{\ \raise.3ex\hbox{$>$\kern-.75em\lower1ex\hbox{$\sim$}}\ }

\title{Remarks on the Warped Deformed Conifold
\footnote{Based on I.R.K.'s talks at the Lisbon School on Superstrings II, 
July 13--17, 2001 and at the Benasque Workshop
``Physics in the Pyrenees: Strings, Branes and Field Theory,''
July 15--27, 2001.}}
\author{Christopher P. Herzog, Igor R. Klebanov and
Peter Ouyang\\ 
Department of Physics, Princeton University\\
Princeton, NJ 08544, USA\\ 
  }
\begin{document}
\setlength{\baselineskip}{16pt}
\begin{titlepage}
\maketitle
\begin{picture}(0,0)(0,0)
\put(325,245){PUPT-2003}
\put(325,260){hep-th/0108101}
\end{picture}
\vspace{-36pt}
\begin{abstract}
We assemble a few remarks on the supergravity solution
of hep-th/0007191, whose UV asymptotic form
was previously found in hep-th/0002159. 
First, by normalizing the R-R fluxes, we
compare the logarithmic flow of couplings in supergravity with that
in field theory, and find exact agreement.
We also write the 3-form field strength $G_3 = F_3 - \tau H_3$
present in the solution in a manifestly $SO(4)$ invariant $(2,1)$
form.
In addition, we discuss various issues related to the chiral symmetry breaking
and wrapped branes.
\end{abstract}
\thispagestyle{empty}
\setcounter{page}{0}
\end{titlepage}

\section{Introduction}

The warped deformed conifold \cite{KS} is a solution of type IIB
supergravity that is dual to a certain ${\cal N}=1$ supersymmetric
$SU(N+M)\times SU(N)$ gauge theory in the limit of strong `t Hooft
coupling. This solution encodes various interesting gauge theory
phenomena in a dual geometrical language, such as the duality cascade
in the UV and chiral symmetry breaking and confinement in the IR.

In these notes we assemble a few remarks on the solution of \cite{KS},
whose UV asymptotic form was previously found in \cite{KT}.  Our
intention is to compromise between presenting a self-contained review
of supergravity conifold solutions and their field theory duals, and
presenting three or four new results that we believe to be of general
interest.  A summary of the new results follows.

First, by normalizing the R-R fluxes, we compare the logarithmic flow
of couplings in supergravity with that in field theory, and find exact
agreement.\footnote{
For ${\cal N}=2$ supersymmetric gauge theories
realized on fractional branes at orbifold singularities,
the agreement of supergravity with field theory $\beta$-functions
was demonstrated in \cite{KN,Bert,Joe}.}  
Next, we show that the 3-form field strength $G_3 = F_3 -
\tau H_3$ present in the solution is an $SO(4)$ invariant $(2,1)$ form
on the deformed conifold.  (It was shown previously that $G_3$ must be
$(2,1)$ in order to preserve SUSY 
\cite{Sav},\cite{Grana}, \cite{Gub}.)  Although
$G_3$ has appeared explicitly in the literature before \cite{Cvet},
the basis in which we write $G_3$ and some other differential forms
important to the supergravity solutions of \cite{KT} and \cite{KS} is
a particularly simple one in which many of the properties of these
forms become completely obvious.

We also discuss various issues related to the chiral symmetry breaking
and wrapped branes.  For example, we develop the gauge field/string
dictionary for this system by showing the correspondence between a
certain wrapped D5-brane in supergravity and domain walls in the field
theory that interpolate between inequivalent vacua.  Finally, in the
style of \cite{uvir}, we discuss various UV/IR relations for the
conifold.

We now review some basic facts about the AdS/CFT correspondence first
because we would like these notes to be self-contained and second
because it is important to understand the normalizations here in view
of the more complicated solutions ahead.  The duality between ${\cal
N}=4$ supersymmetric $SU(N)$ gauge theory and the $AdS_5\times \S^5$
background of type IIB string theory \cite{jthroat,US,EW} is usually
motivated by considering a stack of a large number $N$ of
D3-branes. The SYM theory is the low-energy limit of the gauge theory
on the stack of D3-branes.

On the other hand, the curved 
background produced by the stack is
\be
\label{geom}
ds^2 = h^{-1/2}
\left (- dt^2 +dx_1^2+ dx_2^2+ dx_3^2\right )
+ h^{1/2}
\left ( dr^2 + r^2 d\Omega_5^2 \right )\ ,
\ee
where $d\Omega_5^2$ is the metric of a unit 5-sphere and
\be
h(r) = 1+{L^4\over r^4}\ .
\ee
This 10-dimensional metric may be thought of as a ``warped product''
of the $\R{3,1}$ along the branes and the transverse space $\R{6}$.
Note that the dilaton, $\Phi=0$, is constant, and
the selfdual 5-form field strength is given by
\be \label{firstfive}
F_5 = {\cal F}_5 + \star {\cal F}_5 \ , \qquad  {\cal F}_5
=16 \pi (\alpha')^2 N {\rm vol}(\S^5)
\ .
\ee
The normalization above is dictated by the 
quantization of  D$p$-brane tension which implies
\be \label{quantten}
\int_{\S^{8-p}} \star F_{p+2} = {2\kappa^2 \tau_p N \over g_s}
\ ,
\ee
where
\be
\tau_p = {\sqrt\pi\over \kappa} (4\pi^2\alpha')^{(3-p)/2}
\ee
and $\kappa = 8 \pi^{7/2} g_s \alpha'^2$ is the 10-dimensional
gravitational constant.
In particular, for $p=3$ we have
\be
\int_{\S^5} F_5 = (4\pi^2 \alpha')^2 N
\ ,\ee
which is consistent with (\ref{firstfive})
since the volume of a unit 5-sphere is ${\rm Vol}(\S^5)=\pi^3$.

Note that the 5-form field strength may also be written as
\be
g_s F_5 = d^4 x\wedge d h^{-1} - r^5 {dh\over dr} {\rm vol}(\S^5) \ .
\ee
Then it is not hard to see that
the Einstein equation
$R_{MN} = g_s^2 F_{MPQRS} {F_N}^{PQRS} / 96$ is satisfied.
Since $-r^5 {dh\over dr} = 4 L^4$, we find
by comparing with (\ref{firstfive}) that
\be
L^4 = 4\pi g_s N \alpha'^2\ .
\ee

A related way to determine the scale factor $L$ is to equate the ADM
tension of the supergravity solution with $N$ times the tension of a
single D3-brane \cite{gkp}:
\be
{2\over \kappa^2} L^4 {\rm Vol}(\S^5) = {\sqrt \pi\over \kappa} N
\ ,
\ee
This way we find
\be
L^4 = {\kappa N\over 2\pi^{5/2}} = 4\pi g_s N \alpha'^2\ 
\ee
in agreement with the preceding paragraph.

The radial coordinate $r$ is related to the scale in the dual gauge
theory.  The low-energy limit corresponds to $r\rightarrow 0$.  In
this limit the metric becomes
\be \label{adsmetric}
ds^2 = {L^2 \over z^2} \left( -dt^2 + d\vec{x}^2 + dz^2 \right) +
    L^2 d\Omega_5^2 \ ,
\ee
where $z={L^2\over r}$. This describes the direct product of
5-dimensional Anti-de Sitter space, $AdS_5$, and the 5-dimensional
sphere, ${\bf S}^5$, with equal radii of curvature $L$.

An interesting generalization of the basic AdS/CFT correspondence
\cite{jthroat,US,EW} is found by studying branes at conical
singularities \cite{ks,Kehag,KW,MP}.  Consider a stack of D3-branes
placed at the apex of a Ricci-flat 6-d cone $Y_6$ whose base is a 5-d
Einstein manifold $X_5$. Comparing the metric with the D-brane
description leads one to conjecture that type IIB string theory on
$AdS_5\times X_5$ is dual to the low-energy limit of the world volume
theory on the D3-branes at the singularity. The equality of tensions
now requires \cite{Gubser}
\be \label{radius}
L^4 = {\sqrt \pi \kappa N\over 2 {\rm Vol}(X_5) } = 4\pi g_s N \alpha'^2
{\pi^3\over {\rm Vol}(X_5) } \ ,
\ee
an important normalization formula which we will use in the following
subsection.

The simplest examples of $X_5$ are the orbifolds ${\bf S}^5/\Gamma$
where $\Gamma$ is a discrete subgroup of $SO(6)$ \cite{ks}. In these
cases $X_5$ has the local geometry of a 5-sphere.  The dual gauge
theory is the IR limit of the world volume theory on a stack of $N$
D3-branes placed at the orbifold singularity of $\R{6}/\Gamma$. Such
theories typically involve product gauge groups $SU(N)^k$ coupled to
matter in bifundamental representations \cite{dm}.

Constructions of the dual gauge theories for Einstein manifolds $X_5$
which are not locally equivalent to ${\bf S}^5$ are also possible.
The simplest example is the Romans compactification on $X_5= T^{1,1}=
(SU(2)\times SU(2))/U(1)$ \cite{Romans,KW}.  The dual gauge theory is
the conformal limit of the world volume theory on a stack of $N$
D3-branes placed at the singularity of a Calabi-Yau manifold known as
the conifold \cite{KW}, which is a cone over $T^{1,1}$. Let us explain
this connection in more detail.

\subsection{D3-branes on the Conifold}

The conifold may be described by the
following equation in four complex variables,
\be \label{coni}
\sum_{a=1}^4 z_a^2 = 0
\ .
\ee
Since this equation is invariant under an overall real rescaling of
the coordinates, this space is a cone. Remarkably, the base of this
cone is precisely the space $T^{1,1}$ \cite{cd,KW}. In fact, the
metric on the conifold may be cast in the form \cite{cd}
\be
ds_6^2 = dr^2 + r^2 ds_{T^{1,1}}^2\ ,
\label{conimetric}
\ee
where 
\begin{equation} \label{co}
ds_{T^{1,1}}^2=
{1\over 9} \bigg(d\psi + 
\sum_{i=1}^2 \cos \theta_i d\phi_i\bigg)^2+
{1\over 6} \sum_{i=1}^2 \left(
d\theta_i^2 + {\rm sin}^2\theta_i d\phi_i^2
 \right)
\ .
\end{equation}
is the metric on $T^{1,1}$. Here $\psi$ is an angular coordinate
which  ranges from $0$ to $4\pi$,  while $(\theta_1,\phi_1)$
and $(\theta_2,\phi_2)$ parametrize two ${\bf S}^2$'s in a standard way.
Therefore, this form of the metric shows that
$T^{1,1}$ is an ${\bf S}^1$ bundle over ${\bf S}^2 \times {\bf S}^2$.

Now placing $N$ D3-branes at the apex of the cone we find the metric
\be
\label{newgeom}
ds^2 = \left (1+{L^4\over r^4}\right )^{-1/2}
\left (- dt^2 +dx_1^2+ dx_2^2+ dx_3^2\right )
+ \left (1+{L^4\over r^4}\right )^{1/2}
\left ( dr^2 + r^2 ds_{T^{1,1}}^2 \right )
\ee
whose near-horizon limit is $AdS_5\times T^{1,1}$.
Using the metric (\ref{co}) it is not hard to find that the
volume of $T^{1,1}$ is ${16 \pi^3 \over 27}$ 
\cite{GK}. From (\ref{radius}) it then follows that
\be \label{quantiz}
L^4 = 4\pi g_s N (\alpha')^2 {27\over 16} = 
{27 \kappa N\over 32\pi^{5/2}}
\ .
\ee
The same logic that leads us to the 
${\mathcal N}=4$
version of the AdS/CFT correspondence now shows that
the type IIB string theory on this space should
be dual to the infrared limit of the field theory on $N$ D3-branes
placed at the singularity of the conifold. Since Calabi-Yau spaces
preserve 1/4 of the original supersymmetries 
this should be an ${\cal N}=1$
superconformal field theory. 
This field theory was constructed
in \cite{KW}: it is $SU(N)\times SU(N)$ gauge theory
coupled to two chiral superfields, $A_i$, in the $({\bf N}, \overline{\bf N})$
representation
and two chiral superfields, $B_j$, in the $(\overline{\bf N}, {\bf N})$
representation. 

In order to match the two 
gauge couplings to the moduli of the type IIB
theory on $AdS_5\times T^{1,1}$, one notes that the integrals over the
$\S^2$ of $T^{1,1}$ of the NS-NS and R-R 2-form potentials, $B_2$ and
$C_2$, are moduli. In particular, the two gauge couplings are
determined as follows \cite{KW,MP}:\footnote{Exactly the same relations
apply to the ${\cal N}=2$ supersymmetric $\ZZ_2$ orbifold theory
\cite{ks,Joe}.}
\be
{4\pi^2 \over g_1^2} + {4\pi^2 \over g_2^2} ={\pi\over g_s e^\Phi} 
\ ,
\ee
\be \label{couplediff}
\left [ {4\pi^2 \over g_1^2} - {4\pi^2 \over g_2^2}\right ] 
g_s e^\Phi
= {1\over 2\pi \alpha'}
\left(\int_{\S^2} B_2\right) - \pi \quad ({\rm mod}\ 2\pi)
\ .
\ee
These equations are crucial for relating the SUGRA
background to the field theory $\beta$-functions when the
theory is generalized to $SU(N+M)\times SU(N)$ \cite{KN,KT}. From 
the quantization condition on $H_3$, 
${1\over 2\pi \alpha'}(\int_{\S^2} B_2)$ 
must be a 
periodic variable with period $2\pi$.
This periodicity is crucial for the cascade phenomenon that we discuss
in the next section.

\section{The RG cascade}

The addition of $M$ fractional 3-branes (wrapped D5-branes) at the
singular point changes the gauge group to $SU(N+M)\times SU(N)$.
Let us consider the effect on the dual supergravity background of
adding $M$ wrapped D5-branes.  The D5-branes serve as sources of the
magnetic RR 3-form flux through the $\S^3$ of $T^{1,1}$. Therefore,
the supergravity dual of this field theory involves $M$ units of the
3-form flux, in addition to $N$ units of the 5-form flux:
\be \label{qcond}
{1\over 4\pi^2 \alpha'}\int_{\S^3} F_3 = M\ ,
\qquad\qquad {1\over (4 \pi^2 \alpha')^2 }\int_{T^{1,1}} F_5 = N
\ .
\ee
The coefficients above follow from the quantization rule
(\ref{quantten}).  The warped conifold solution with such fluxes was
constructed in \cite{KT}.

It will be useful to employ
the following basis of 1-forms on the compact space
\cite{MT}:
\bea \label{fbasis}
g^1 = {e^1-e^3\over\sqrt 2}\ ,\qquad
g^2 = {e^2-e^4\over\sqrt 2}\ , \nonumber \\
g^3 = {e^1+e^3\over\sqrt 2}\ ,\qquad
g^4 = {e^2+ e^4\over\sqrt 2}\ , \nonumber \\
g^5 = e^5\ ,
\eea
where
\begin{eqnarray}
e^1\equiv - \sin\theta_1 d\phi_1 \ ,\qquad
e^2\equiv d\theta_1\ , \nonumber \\
e^3\equiv \cos\psi\sin\theta_2 d\phi_2-\sin\psi d\theta_2\ , \nonumber\\
e^4\equiv \sin\psi\sin\theta_2 d\phi_2+\cos\psi d\theta_2\ , \nonumber \\
e^5\equiv d\psi + \cos\theta_1 d\phi_1+ \cos\theta_2 d\phi_2 \ .
\end{eqnarray}
In terms of this basis, the Einstein metric on $T^{1,1}$ assumes the
form
\be
ds^2_{T^{1,1}}= {1\over 9} (g^5)^2 + {1\over 6}\sum_{i=1}^4 (g^i)^2
\ .
\ee

Keeping track of the normalization factors, in order
to be consistent with the quantization conditions (\ref{qcond}),
\be \label{closedf}
F_3 = {M\alpha'\over 2}\omega_3\ ,  \qquad\qquad 
B_2 = {3 g_s M \alpha'\over 2}\omega_2 \ln (r/r_0)
\ ,
\ee
\be
H_3 = dB_2 = {3 g_s M \alpha' \over 2r} dr\wedge \omega_2\ ,
\label{hthree}
\ee
where 
\be
\omega_2 = {1\over 2}(g^1\wedge g^2 + g^3 \wedge g^4)= 
{1\over 2} (\sin\theta_1 d\theta_1 \wedge d\phi_1- 
\sin\theta_2 d\theta_2 \wedge d\phi_2 )\ ,
\label{omegato}
\ee
\be
\omega_3 = {1\over 2} g^5\wedge (g^1\wedge g^2 + g^3 \wedge g^4)\ .
\label{omegathr}
\ee
In Appendix A we show that
\be
\int_{\S^2} \omega_2 = 4 \pi\ , \qquad
\int_{\S^3} \omega_3 = 8 \pi^2 \ 
\label{intforms}
\ee
where the $S^2$ is parametrized by $\psi=0$, $\theta_1=\theta_2$ and
$\phi_1=-\phi_2$, and the $S^3$ by $\theta_2=\phi_2=0$.  As a result,
the quantization condition for RR 3-form flux is obeyed.

Both $\omega_2$ and $\omega_3$ are closed.
Note also that
\be \label{duality}
g_s \star_6 F_3 = H_3\ ,\qquad g_s F_3 =  -\star_6 H_3\ ,
\ee
where $\star_6$ is the Hodge dual with respect to the metric
$ds_6^2$. Thus, the complex 3-form $G_3$
satisfies the self-duality condition
\be
\star_6 G_3 = i G_3\ , \qquad\qquad G_3 = F_3 - {i\over g_s} H_3\ .
\ee
Note that the self-duality fixes the relative factor of 3 in 
(\ref{closedf}) (see (\ref{conimetric}), (\ref{co})). We will see that
this geometrical factor is crucial for reproducing the well-known
factor of 3 in the ${\cal N}=1$ beta functions.

It follows from (\ref{duality}) that
\be \label{dilcons}
g_s^2 F_3^2 = H_3^2
\ ,
\ee
which implies that the dilaton is constant, $\Phi=0$.
Since $F_{3\mu\nu\lambda} H_3^{\mu\nu\lambda} =0$, the RR scalar
vanishes as well. 

The 10-d metric found in \cite{KT} has the structure of a
``warped product'' of $\R{3,1}$ and the conifold:
\be \label{fulsol}
ds^2_{10} =   h^{-1/2}(r)   dx_n dx_n 
 +  h^{1/2} (r)  (dr^2 + r^2 ds^2_{T^{1,1}} )\ .
\ee
The solution 
for the warp factor $h$ may be determined
from the trace of the Einstein equation:
\be
R = {1\over 24} (H_3^2 + g_s^2 F_3^2) = {1\over 12} H_3^2\ .
\ee
This implies
\be
- h^{-3/2} {1\over r^5} {d\over dr} (r^5 h') ={1\over 6} H_3^2  
\ .
\ee 
Integrating this differential equation, we find that 
\be \label{nonharm}
h(r) = {27 \pi (\alpha')^2 [g_s N + 
a (g_s M)^2 \ln (r/r_0) + a (g_s M)^2/4]
\over 4 r^4}
\ee
with $a=3/(2\pi)$.

An important feature of this background
is that $\tilde F_5$ acquires a radial dependence \cite{KT}.
This is because
\be
\tilde F_5 = F_5 + B_2\wedge F_3\ , \qquad F_5 = d C_4\ ,
\ee
and $\omega_2\wedge \omega_3 = 54 {\rm vol}(T^{1,1})$.
Thus, we may write
\be
\tilde F_5 = {\cal F}_5 + \star {\cal F}_5 \ , \qquad  {\cal F}_5
= 27 \pi \alpha'^2 N_{eff} (r) {\rm vol}({\rm T}^{1,1})
\ ,
\ee
and 
\be
N_{eff} (r) = N + {3\over 2 \pi} g_s M^2 \ln (r/r_0)
\ .
\label{Neff}
\ee
The novel phenomenon in this solution is that the 5-form flux present
at the UV scale $r=r_0$ may completely disappear by the time we
reach a scale where $N_{eff} = 0$.
The non-conservation of the flux is due to the type IIB SUGRA equation
\be \label{transg}
d \tilde F_5 = H_3 \wedge F_3
\ .
\ee
A related fact is that 
$\int_{\S^2} B_2$ is 
no longer a periodic variable in the SUGRA solution
once the $M$ fractional branes are introduced:
as the $B_2$ flux goes through a period, $N_{eff} (r) \rightarrow 
N_{eff} (r) - M$
which has the effect of decreasing the 5-form flux by $M$ units.
Note from (\ref{Neff}) that for a single cascade step 
$N_{eff} (r) \rightarrow N_{eff} (r) - M$
the radius changes by a factor
$r_2 / r_1 = \exp(-2\pi / 3 g_s M)$, agreeing with a result of
\cite{GKP}.

Due to the non-vanishing RHS of (\ref{transg}),
${1\over (4 \pi^2 \alpha')^2 }\int_{T^{1,1}} \tilde F_5$
is not quantized. We may identify this quantity with $N_{eff}$
defining the gauge group $SU(N_{eff}+M)\times SU(N_{eff})$
only at special radii $r_k= r_0 \exp(-2\pi k/ 3 g_s M)$ where
$k$ is an integer. Thus, $N_{eff}= N - kM$.
Furthermore, we believe that the continuous 
logarithmic variation of $N_{eff} (r)$ 
is related to continuous reduction in the number of degrees of
freedom as the theory flows to the IR. Some support for this claim
comes from studying the high-temperature phase of this theory
using black holes embedded into asymptotic KT grometry 
\cite{finitetemp}. The effective number of degrees of freedom
computed from the Bekenstein--Hawking entropy grows logarithmically with
the temperature, in agreement with (\ref{Neff}).

The metric (\ref{fulsol}) 
has a naked singularity at $r=r_s$ where $h(r_s)=0$.
Writing
\be \label{UVs}
h(r) = {L^4\over r^4} \ln (r/r_s)\ , \qquad 
L^2 = {9  g_s M \alpha'\over 2\sqrt 2}
\ ,
\ee
we find a purely logarithmic RG cascade:
\be \label{dualmet}
ds^2 = {r^2\over L^2 \sqrt{\ln (r/r_s)} } dx_n dx_n
+ { L^2 \sqrt{\ln (r/r_s)}\over r^2} dr^2 + L^2 \sqrt{\ln (r/r_s)} 
ds^2_{T^{1,1}}\ .
\ee
Since $T^{1,1}$ expands
slowly toward large $r$, the curvatures decrease there so
that corrections to the SUGRA become negligible. Therefore, 
even if $g_s M$ is very small,
this SUGRA solution is reliable for sufficiently 
large radii where $g_s N_{eff} (r)\gg 1$.
In this regime the separation
between the cascade steps is very large, so that the SUGRA
calculation of the $\beta$-functions may be compared with
$SU(N_{eff}+ M)\times SU(N_{eff})$ gauge theory.
We will work near $r=r_0$ where $N_{eff}$ may be replaced by $N$.

\subsection{Matching of the $\beta$-functions}

In gauge/gravity duality the 5-dimensional radial coordinate
defines the RG scale of the dual gauge theory 
\cite{jthroat,US,EW,holobound,uvir}.
There are different ways of establishiing the precise relation.
The simplest one is to identify the field theory energy scale
$\Lambda$
with the energy of a stretched string ending on a probe brane
positioned at radius $r$. For all metrics of the form (\ref{fulsol})
this gives
\be
\Lambda\sim r
\ .
\ee
In this section we adopt this UV/IR relation, which typically
corresponds to the Wilsonian renormalization group.  We will discuss
UV/IR relations in more detail in Section 5.

Now we are ready to interpret the solution of \cite{KT} in terms
of RG flow in the dual $SU(N+M)\times SU(N)$ gauge theory.
The constancy of the dilaton translates into the vanishing
of the $\beta$-function for
$ {8\pi^2\over g_1^2} + {8\pi^2\over g_2^2}$.
Substituting the solution for $B_2$ into (\ref{couplediff})
we find 
\be \label{betres}
{8\pi^2\over g_1^2} - {8\pi^2\over g_2^2} = 6 M \ln (r/r_s) + {\rm
const}
\ .
\ee
Since $\ln (r/r_s) = \ln (\Lambda/\mu)$,  (\ref{betres})
implies a logarithmic running of
${1\over g_1^2} - {1\over g_2^2}$ in the $SU(N+M)\times SU(N)$ gauge theory.
As we mentioned earlier, this SUGRA result is reliable
for any value of $g_s M$ provided that
$g_s N \gg 1$. We may consider $g_s M \ll 1$ so that the cascade
jumps are well-separated.

Let us compare with the Shifman--Vainshtein
$\beta$-functions \cite{NSVZ}:\footnote{
These expressions for the $\beta$-functions
differ from the standard NSVZ form 
\cite{NSVZinstanton} by a factor of $1/(1 - g^2 N_c / 8\pi^2)$.
The difference comes from the choice of
normalization of the vector superfields.
We choose the normalization so that the relevant kinetic term in the
field theory action is $\frac{1}{2g^2}\int d^4x d^2\theta
Tr(W^{\alpha} W_{\alpha})+$ h.c.;
this choice is dictated by the form of the supergravity action and
differs from the canonical normalization by a factor of $1/g^2$.
With this convention the additional factor in the $\beta$-function does 
not appear.
A nice review of the derivation of the exact
$\beta$-functions is in \cite{susynotes}.}
\beq \label{SVexact} 
&{d\over d {\rm log} (\Lambda/\mu)}
{8 \pi^2\over g_1^2} & = 3(N +M) - 2N (1- \gamma)\ ,\\
&{d\over d {\rm log} (\Lambda/\mu)}
{8 \pi^2 \over g_2^2} & = 3N - 2(N+M) (1-  \gamma)\ ,
\eeq
where $\gamma$ is the anomalous dimension 
of operators ${\rm Tr} A_i B_j$.
The conformal invariance of the field theory for $M=0$,
and symmetry under $M\rightarrow -M$,
require that 
$\gamma =-{1\over 2} + O[(M/N)^{2n}]$ where $n$
is a positive integer \cite{KS}.  
Taking the difference of the two equations in (\ref{SVexact}) we then
find
\be \label{betafun} 
{8\pi^2 \over g_1^2} - {8\pi^2 \over g_2^2} =
 M \ln (\Lambda/\mu) [ 3 + 2 
(1-\gamma)]= 6 M \ln (\Lambda/\mu) (1+ O[(M/N)^{2n}])\ .
\ee
Remarkably, the coefficient $6M$ is in {\it exact} agreement
with the result
(\ref{betres}) found on the SUGRA side.  
This consitutes a geometrical explanation of a field theory $\beta$-function,
including its normalization. 

We may also 
trace the jumps in the rank of the gauge group to a well-known
phenomenon in the dual ${\cal N}=1$ field theory, namely, Seiberg
duality \cite{NAD}. The essential observation is that $1/g_1^2$ and
$1/g_2^2$ flow in opposite directions and, according to
(\ref{SVexact}), there is a scale where the $SU(N+M)$ coupling, $g_1$,
diverges. To continue past this infinite coupling, we perform a ${\cal
N}=1$ duality transformation on this gauge group factor.  The
$SU(N+M)$ gauge factor has $2N$ flavors in the fundamental
representation.  Under a Seiberg duality transformation, this becomes
an $SU(2N-[N+M]) = SU(N-M)$ gauge group. 
Thus we obtain an $SU(N)\times SU(N-M)$
theory which resembles closely the theory we started with \cite{KS}.

As the theory flows to the IR, the cascade must stop, however, because
negative $N$ is physically nonsensical. Thus, we should not
be able to continue the solution (\ref{dualmet}) to 
the region where $N_{eff}$ is negative.
To summarize,
the fact that the solution of \cite{KT} is singular
tells us that it has to be modified in the IR.

\section{Deformation of the Conifold}

It was shown in \cite{KS}
that, to remove the naked singularity found in
\cite{KT} the conifold (\ref{coni}) should be replaced by the deformed
conifold
\begin{equation} \label{dconifold}
\sum_{i=1}^4 z_i^2 = 
\varepsilon^2\ ,
\end{equation}
in which the singularity of the conifold is removed
through the blowing-up of the  $\S^3$ of $T^{1,1}$.
We now review the deformed conifold in order to be able 
to normalize properly the field strengths and to prepare for a discussion
of a new and simple $SO(4)$ invariant way of writing the field strengths.
The 10-d metric of \cite{KS} 
takes the following form:
\be \label{specans}
ds^2_{10} =   h^{-1/2}(\tau)   dx_n dx_n 
 +  h^{1/2}(\tau) ds_6^2 \ ,
\ee
where $ds_6^2$ is the metric of the deformed conifold (\ref{metricd}).
This is the same type of ``D-brane'' ansatz as (\ref{fulsol}), but with the
conifold replaced by the deformed conifold as 
the transverse space.
 
The metric of the deformed conifold was discussed in some detail in
\cite{cd,MT,Ohta}. It is diagonal in the basis (\ref{fbasis}):
\bea \label{metricd}
ds_6^2 = {1\over 2}\varepsilon^{4/3} K(\tau)
\Bigg[ {1\over 3 K^3(\tau)} (d\tau^2 + (g^5)^2) 
 + 
\cosh^2 \left({\tau\over 2}\right) [(g^3)^2 + (g^4)^2]\nonumber \\
+ \sinh^2 \left({\tau\over 2}\right)  [(g^1)^2 + (g^2)^2] \Bigg]
\ ,
\eea
where
\be
K(\tau)= { (\sinh (2\tau) - 2\tau)^{1/3}\over 2^{1/3} \sinh \tau}
\ .
\ee
{}For large $\tau$ we may introduce another radial coordinate $r$ via
\be \label{changeofc}
r^2 = {3\over 2^{5/3}} \varepsilon^{4/3} e^{2\tau/3}\ ,
\ee
and in terms of this radial coordinate 
$ ds_6^2 \rightarrow dr^2 + r^2 ds^2_{T^{1,1}}$.

At $\tau=0$ the angular metric degenerates into
\be 
d\Omega_3^2= {1\over 2} \varepsilon^{4/3} (2/3)^{1/3}
[ {1\over 2} (g^5)^2 + (g^3)^2 + (g^4)^2 ]
\ ,
\ee
which is the metric of a round $\S^3$ \cite{cd,MT}.
The additional two directions, corresponding to the $\S^2$ fibered
over the $\S^3$, shrink as
\be {1\over 8} \varepsilon^{4/3} (2/3)^{1/3}
\tau^2 [(g^1)^2 + (g^2)^2]
\ .\ee

The simplest ansatz for the 2-form fields is
\bea
F_3 = {M\alpha'\over 2} \left \{g^5\wedge g^3\wedge g^4 + d [ F(\tau) 
(g^1\wedge g^3 + g^2\wedge g^4)]\right \} \nonumber \\
= {M\alpha'\over 2} \left \{g^5\wedge g^3\wedge g^4 (1- F)
+ g^5\wedge g^1\wedge g^2 F + F' d\tau\wedge
(g^1\wedge g^3 + g^2\wedge g^4) \right \}\ ,
\eea
with $F(0) = 0$ and $F(\infty)=1/2$, and
\be
B_2 = {g_s M \alpha'\over 2} [f(\tau) g^1\wedge g^2
+  k(\tau) g^3\wedge g^4 ]\ ,
\ee
\be
H_3 = dB_2 = {g_s M \alpha'\over 2} [d\tau\wedge (f' g^1\wedge g^2
+  k' g^3\wedge g^4) + {1\over 2} (k-f) 
g^5\wedge (g^1\wedge g^3 + g^2\wedge g^4) ]\ .
\ee

As before, the self-dual 5-form field strength may be 
decomposed as $\tilde F_5 = {\cal F}_5 + \star {\cal F}_5$. We
have
\be
{\cal F}_5 = B_2\wedge F_3 = {g_s M^2 (\alpha')^2\over 4} \ell(\tau)
g^1\wedge g^2\wedge g^3\wedge g^4\wedge g^5\ ,
\ee
where
\be
\ell = f(1-F) + k F\ ,
\ee
and
\be
\star {\cal F}_5 = 4 g_s M^2 (\alpha')^2 \varepsilon^{-8/3}
dx^0\wedge dx^1\wedge dx^2\wedge dx^3
\wedge d\tau {\ell(\tau)\over K^2 h^2 \sinh^2 (\tau)}\ .
\ee


\subsection{The First-Order Equations and Their Solution}

In searching for BPS saturated
supergravity backgrounds, the second order equations should be replaced by
a system of first-order ones. 
Luckily, this is possible for our ansatz \cite{KS}:
\bea \label{firstorder}
f' &=& (1-F) \tanh^2 (\tau/2)\ , \nonumber \\
k' &=& F \coth^2 (\tau/2)\ , \nonumber \\
F' &=& {1\over 2} (k-f)\ , 
\eea
and
\be \label{firstgrav}
h' = - \alpha {f(1-F) + kF\over K^2 (\tau) \sinh^2 \tau}
\ ,
\ee
where 
\be
\alpha =4 (g_s M \alpha')^2
\varepsilon^{-8/3}\ .
\ee
These equations follow from a superpotential for the effective radial
problem \cite{PT}.
Once we have the solutions to these differential equations, 
we can check that the large $\tau$ limit of 
the properly normalized $B_2$, $F_3$ and 
$F_5$ field strengths agree with their simpler counterparts
of section 2.  Also, we can understand precisely the
large and small $\tau$ behavior of the warp factor $h(\tau)$.

Note that the first three of these equations,
(\ref{firstorder}), form a closed system and need to be
solved first. 
In fact, these equations imply the self-duality of the
complex 3-form with respect to the metric of the
deformed conifold: $\star_6 G_3 = i G_3$.
The solution is
\bea 
F(\tau) &=& {\sinh \tau -\tau\over 2\sinh\tau}\ ,
\nonumber \\ 
f(\tau) &=& {\tau\coth\tau - 1\over 2\sinh\tau}(\cosh\tau-1) \ ,
\nonumber \\ 
k(\tau) &=& {\tau\coth\tau - 1\over 2\sinh\tau}(\cosh\tau+1)
\ .
\eea

Now that we have solved for the 3-forms on the deformed conifold,
the warp factor may be determined by integrating
(\ref{firstgrav}).
First we note that
\be
\ell(\tau) = f(1-F) + kF =  {\tau\coth\tau - 1\over 4\sinh^2\tau}
(\sinh 2\tau-2\tau)
\ .\ee
This behaves as $\tau^3$ for small $\tau$.
For large $\tau$ we impose, as usual, the boundary condition that
$h$ vanishes. The resulting integral expression for $h$ is
\be \label{intsol}
h(\tau) = \alpha { 2^{2/3}\over 4} I(\tau) = 
(g_s M\alpha')^2 2^{2/3} \varepsilon^{-8/3} I(\tau)\ ,
\ee
where
\be
I(\tau) \equiv
\int_\tau^\infty d x {x\coth x-1\over \sinh^2 x} (\sinh (2x) - 2x)^{1/3}
\ .
\ee 
We have not succeeded in evaluating this integral in terms of elementary
or well-known special functions, but it is not hard to see that 
\be
I(\tau\to 0) \to a_0 + O(\tau^2) \ ; \qquad 
\ I(\tau\to\infty)\to 3 \cdot 2^{-1/3} 
\left (\tau - {1\over 4} \right ) e^{-4\tau/3}
\ ,\ee
where $a_0\approx 0.71805$.
This $I(\tau)$ is nonsingular at the tip of the deformed conifold and, from
\eref{changeofc}, matches the form of the large-$\tau$ solution
\eref{UVs}.  The small $\tau$ behavior follows from the
convergence of the integral \eref{intsol}, while at large
$\tau$ the integrand becomes $\sim xe^{-4x/3}$.

Thus,
for small $\tau$ the ten-dimensional geometry is 
approximately $\R{3,1}$ times the 
deformed conifold:
\bea \label{apex}
ds_{10}^2  \rightarrow  &{ \varepsilon^{4/3}\over 
2^{1/3} a_0^{1/2} g_s M\alpha'} dx_n dx_n 
+   a_0^{1/2} 6^{-1/3} (g_s M\alpha')
\bigg \{ {1\over 2} d\tau^2  + {1\over 2} (g^5)^2
+ (g^3)^2 + (g^4)^2   \nonumber \\  & + {1\over 4}\tau^2
[(g^1)^2 + (g^2)^2] \bigg \}
\ .
\eea
Note that we have suppressed the ${\mathcal O}(\tau^2)$ 
corrections for all but the $(g^1)^2$ and $(g^2)^2$ components
of the metric.
This metric will be useful in section 4 where we
investigate various infrared phenomenon of the gauge theory.

Very importantly, for large $g_s M$
the curvatures found in our solution are small
everywhere.
 This is true even far in the IR, since
the radius-squared of the $\S^3$ at $\tau=0$ is of order 
$g_s M $ in  string units. This
is the `t Hooft coupling of the gauge theory found far in the IR.
As long as this is large, the curvatures are small and the SUGRA
approximation is reliable.

\subsection{$SO(4)$ invariant expressions for the 3-forms}

In \cite{Grana,Gub} it was shown that the warped background
of the previous section preserves ${\cal N}=1$ SUSY if and only
if
$G_3$ is a $(2,1)$ form on the CY space.
Perhaps the easiest way to see the supersymmetry of 
the deformed conifold solution is through a T-duality.
Performing a T-duality along one of the longitudinal directions,
and lifting the result to M-theory maps our background to a Becker-Becker
solution supported by a $G_4$ which is a $(2,2)$ form on
$T^2\times {\rm CY}$. G-flux of this type indeed produces a supersymmetric
background \cite{Becker}.

While writing $G_3$ in terms of the angular 1-forms
$g^i$ is convenient for some purposes, the $(2,1)$ nature of the form
is not manifest. That $G_3$ is indeed $(2,1)$ 
was demonstrated in \cite{Cvet} with the help
of a holomorphic basis.
Below we write the $G_3$ found in 
\cite{KS} in terms of the obvious 1-forms
on the deformed conifold: $dz^i$ and $d\bar z^i$, $i=1,2,3,4$:
\bea
G_3 = \frac{M \alpha'}{2\varepsilon^6 \sinh^4 \tau}
\bigg\{
\frac{\sinh (2\tau)-2\tau}{\sinh \tau}
(\epsilon_{ijkl} \, z_i \bar{z}_j \, dz_k \wedge d\bar{z}_l)
\wedge (\bar{z}_m \, dz_m) \nonumber \\
+ 2 ( 1 - \tau\coth\tau )
(\epsilon_{ijkl} \, z_i \bar{z}_j \, dz_k \wedge dz_l)
\wedge (z_m \, d\bar{z}_m)\bigg \}  .
\eea
We also note that the NS-NS 2-form potential is an
$SO(4)$ invariant $(1,1)$ form:
\be
B_2 = \frac{i g_s M \alpha'}{2 \varepsilon^4} 
\frac{\tau \coth \tau - 1}{\sinh^2 \tau}
\epsilon_{ijkl} \, z_i \bar{z}_j \, dz_k \wedge d\bar{z}_l
 \ .
\ee
The derivation of these formulae is given in Appendix B.
Our expressions for the gauge fields are manifestly 
$SO(4)$ invariant, and so is the metric. This completes the proof
of $SO(4)$ invariance of the KS solution.

\section{Infrared Physics}

We have now seen that the deformation of the conifold allows the
solution to be non-singular. 
In the following sections we point out some interesting features of the SUGRA
background we have found and show how they realize the expected
phenomena in the dual field theory.  In particular, we will now
demonstrate that there is confinement;
that the theory has glueballs and baryons whose mass scale emerges
through a dimensional transmutation;
that there is a gluino condensate that breaks the $\ZZ_{2M}$ chiral
symmetry down to $\ZZ_2$, and correspondingly there are domain walls
separating inequivalent vacua.
Other stringy approaches to infrared phenomena in ${\cal N}=1$
SYM theory have recently appeared in \cite{MN,Vafa}.

\subsection{Dimensional Transmutation and Confinement}

The resolution of the naked singularity via the deformation of the
conifold is a supergravity realization of the dimensional transmutation.
While the singular conifold has no dimensionful parameter, we saw that
turning on the R-R 3-form flux produces 
the logarithmic warping of the 
KT solution. 
The scale necessary to define the
logarithm transmutes into the
the parameter $\varepsilon$ that determines the deformation of the
conifold. From (\ref{changeofc}) we see that $\varepsilon^{2/3}$
has dimensions of length and that
\be 
\tau = 3 \ln (r/\varepsilon^{2/3}) + {\rm const}
\ .
\ee
Thus, the scale $r_s$ entering the UV solution 
(\ref{UVs}) should be identified with
$\varepsilon^{2/3}$. 
On the other hand, the form of the IR metric (\ref{apex})
makes it clear that the dynamically generated
4-d mass scale, which sets the tension of the
confining flux tubes, is
\be
{\varepsilon^{2/3}\over \alpha' \sqrt{ g_s M}
}
\ .
\label{IRscale}
\ee

The reason the theory is confining is 
that in the metric for small $\tau$ (\ref{apex}) the function
multiplying $dx_n dx_n$ approaches a constant. This should be
contrasted with the $AdS_5$ metric where this function vanishes at the
horizon, or with the singular metric of \cite{KT} where it blows up.
Consider a Wilson contour positioned at fixed $\tau$, and
calculate the expectation value of the Wilson loop using the
prescription \cite{Juan,Rey}. The minimal area surface bounded by the
contour bends towards smaller $\tau$. If the contour has a very large
area $A$, then most of the minimal surface will drift down into the
region near $\tau=0$. From the fact that the coefficient of $dx_n
dx_n$ is finite at $\tau=0$, we find that a fundamental string with
this surface will have a finite tension, and so the resulting Wilson
loop satisfies the area law.  A simple
estimate shows that the string tension scales as 
\be 
T_s \sim
{\varepsilon^{4/3}\over (\alpha')^2 g_s M } \ .  
\ee

The masses of glueball and Kaluza-Klein (KK) states scale as
\be
m_{glueball} \sim m_{KK} \sim {\varepsilon^{2/3}\over 
g_s M \alpha' }
\ .
\ee
Comparing with the string tension, we see that
\be 
T_s \sim g_s M (m_{glueball} )^2
\ .
\ee

Due to the deformation, the full SUGRA background has a finite
3-cycle.  We may interpret various branes wrapped over this 3-cycle in
terms of the gauge theory. Note that the 3-cycle has the minimal
volume near $\tau=0$, hence all the wrapped branes will be localized
there. A
wrapped D3-brane plays the role of a baryon vertex which ties
together $M$ fundamental strings.  Note that for $M=0$ the D3-brane
wrapped on the $\S^3$ gave a dibaryon \cite{GK}; the connection between
these two objects becomes clearer when one notes that for $M>0$ the
dibaryon has $M$ uncontracted indices, and therefore joins $M$
external charges. Studying a probe D3-brane in the background of our solution
show that the mass of the baryon scales as
\be
M_b \sim M {\varepsilon^{2/3}\over \alpha'}
\ .
\ee  

\subsection{Chiral Symmetry Breaking and Gluino Condensation}

Our $SU(N+M) \times SU(N)$ field theory has an anomaly-free
$\ZZ_{2M}$ R-symmetry at all scales. The UV (large $\tau$) 
limit of our metric,
which coincides with the solution found in \cite{KT}, has a $U(1)$
R-symmetry associated with the rotations of the angular coordinate
$\psi$. However, the background value of the R-R 2-form $C_2$
does not have this continuous symmetry. Although $F_3= dC_2$
given in (\ref{closedf}) is $U(1)$ symmetric,
there is no smooth global expression
for $C_2$. Locally, we may write for large $\tau$,
\be \label{orig}
C_2\rightarrow {M\alpha'\over 2}\psi \omega_2 
\ .
\ee
Under $\psi\rightarrow \psi+\epsilon$,
\be
C_2 \rightarrow C_2 + {M\alpha'\over 2} \epsilon \omega_2
\ .\ee
This modification of the asymptotic value of $C_2$ is dual to the
appearance of opposite $\theta$-angles for the two gauge groups,
which is a manifestation of the anomaly in the $U(1)$ R-symmetry
\cite{Ed,MN}.

Let us show that only the discrete shifts
\be \label{discretesh}
\psi \rightarrow \psi +{2 \pi k\over M}\ , \qquad k=1, 2, \ldots, M
\ee
are symmetries of the UV theory.
To this end we may consider domain walls made of $k$ D5-branes wrapped
over the finite-sized $\S^3$ at $\tau=0$, with remaining directions
parallel to $\R{3,1}$. Such a domain wall is obviously a stable
object in the KS background and crossing it takes us from one ground state
of the theory to another.
Indeed, the wrapped D5-brane produces a discontinuity
in $\int_B F_3$, where $B$ is the cycle dual to the $\S^3$.
If to the left of the domain wall $\int_B F_3=0$, as in the basic solution
derived in the preceding sections, then to the right of the domain
wall
\be
\int_B F_3 = 4\pi^2 \alpha' k
\ ,\ee
as follows from the quantization of the D5-brane charge.
The B-cycle is bounded by a 2-sphere at $\tau=\infty$, hence
$\int_B F_3= \int_{\S^2} \Delta C_2$. Therefore
from (\ref{intforms}) it is clear that 
to the right of the wall
\be
\Delta C_2\rightarrow \pi \alpha' k \omega_2
\ee
for large $\tau$. 
This change in $C_2$ is produced by the $\ZZ_{2M}$
transformation (\ref{discretesh}) on the original field configuration
(\ref{orig}).

Recalling that $\psi$ ranges from $0$ to $4\pi$, we see that the full
solution, which depends on $\psi$ through 
$\cos \psi$ and $\sin \psi$, has the $\ZZ_2$ symmetry generated by 
$\psi \rightarrow \psi + 2\pi$. Therefore, a domain wall made
of $M$ D5-branes returns the solution to itself. There are exactly $M$
different discrete orientations of the solution, corresponding to
breaking of the $\ZZ_{2M}$ UV symmetry through the IR effects.
The domain walls constructed out of the wrapped D5-branes separate
these inequivalent vacua.
As we expect, flux tubes can
end on these domain walls \cite{dvalishif}, 
and baryons can dissolve in them. By studying a probe D5-brane 
in the metric, we find that the domain wall tension is
\be
T_{wall} \sim {1\over g_s} {\varepsilon^2\over (\alpha')^{3}}\ .
\ee

In supersymmetric gluodynamics
the breaking of chiral symmetry is associated with the gluino condensate
$\vev{\lambda\lambda}$. A holographic calculation of the
condensate was carried out by Loewy and Sonnenschein in \cite{SL} 
(see also \cite{Bigazzi} for previous work on gluino condensation 
in conifold theories.) 
They looked for the deviation of the complex 2-form field
$C_2 - {i\over g_s} B_2$ from its asymptotic large $\tau$ form
that enters the KT solution:
\bea
\delta \left (C_2 - {i\over g_s} B_2 \right)  & \sim
{M\alpha'\over 4} \tau e^{-\tau} [g_1\wedge g_3 + g_2\wedge g_4 -
i( g_1\wedge g_2 - g_3\wedge g_4)]
\nonumber \\ & 
\sim {M\alpha' \varepsilon^2\over r^3}
\ln(r/\varepsilon^{2/3}) e^{i\psi} (d \theta_1 - i\sin \theta_1 d\phi_1)
\wedge  (d \theta_2 - i\sin \theta_2 d\phi_2)
\ .
\eea
In a space-time that approaches $AdS_5$ a perturbation that scales as
$r^{-3}$ corresponds to the expectation value of a dimension 3 operator.
The presence of an extra $\ln(r/\varepsilon^{2/3})$ factor is presumably
due to the fact that the asymptotic KT metric differs from $AdS_5$
by such logarithmic factors. From the angular dependence of the
perturbation we see that the dual operator is $SU(2)\times SU(2)$
invariant and carries R-charge 1. 
These are precisely the properties of $\lambda\lambda$.
Thus, the holographic calculation tells us that
\be
\vev{\lambda\lambda} \sim M {\varepsilon^2\over (\alpha')^3}
\ .
\ee
Thus, the parameter $\varepsilon^2$ which enters the deformed 
conifold equation has a dual interpretation as the gluino 
condensate.\footnote{
It would be nice to understand the relative factor of $g_s M$ 
between $T_{wall}$ and 
$\langle \lambda \lambda \rangle$.
}

\section{UV/IR Relations and the RG Flow}

In this section we investigate some of the consequences of
compactifying the conifold. If the cascade is embedded inside a
compact manifold, as in \cite{GKP}, then the radius $\tau$ is
effectively cut off at some large $\tau_c$.  The radial cutoff is a
scale in the theory, which becomes an ultraviolet cutoff in the
boundary gauge theory.  The precise relation of these distance and
energy scales depends in general on the physics one is investigating.
We are aware of three schemes for relating the two scales: first, one
could consider the energy of a string stretched from the tip of the
conifold to the regularized boundary as a function of $\tau_c$
\cite{jthroat}.  Second, one can try to think about the warp factor
$h(\tau)$ as a redshift factor which relates the energy of a probe in
the bulk of $AdS$ space to its apparent energy as seen by an observer
on the boundary.  Third, one can consider the equations of motion for
supergravity probes; this is sometimes called the holographic scheme
\cite{holobound}.  In conformal backgrounds, the various
distance/energy relations differ only by their normalization; for
$AdS_5\times S^5$, $E \propto r$ in all three schemes.  However, in
non-conformal backgrounds the distance/energy relations can have
different functional forms \cite{uvir}, and we will see that this is
the case for the KT and KS solutions.

One prescription for relating distance and energy scales comes from
considering the energy of a string stretched from 
some fixed $\tau$ to the cutoff
radius $\tau_c$, where it is stabilized by an external force --
by a probe D-brane at $\tau_c$, for example.  The energy of such a string is
proportional to its worldvolume per unit time:
\begin{equation}
   E \sim \int^{\tau_c} \sqrt{g_{tt}g_{\tau \tau}} d\tau \sim e^{\tau_c/3} 
      + \textrm{const}   \sim r_c \ .
\end{equation}
The energy of this string corresponds to the linearly divergent
self-energy of a quark in the boundary gauge theory, and the radial
cutoff of the geometry regulates the divergence.  Decreasing the
radial cutoff removes high energy string modes and thus corresponds to
integrating out high energy gauge theory modes in the Wilsonian sense.
An appealing feature of this prescription is that $\ln
\frac{\Lambda}{\mu} \sim \ln(r_c) +$ const, so that the difference of
the couplings as predicted by supergravity agrees exactly with the
gauge theory expectation, with no additional ln ln terms.

An alternative prescription is to interpret the warp factor $h(\tau)$
as a redshift factor.  An object with energy $E_{\tau}$ at radial
position $\tau$ will appear to an observer at $\tau'$ to have energy
$E_{\tau'}$, where the energies are related by $E_{\tau}h^{1/4}(\tau)=
E_{\tau'}h^{1/4}(\tau')$.  In terms of the renormalization scale, the
distance/energy relation becomes
\begin{equation} \label{gravdim}                   
\Lambda \sim \mu \, I(\tau_c)^{-1/4} \sim \mu \, [ \tau_c^{-1/4} e^{\tau_c/3}  \ldots] \ .
\end{equation}
This redshift relation introduces corrections to (\ref{betafun}) of
the form $\ln \ln (\Lambda/\mu)$.  They have the same form as
corrections to the flow due to two-loop
$\beta$-functions.

We can derive a third distance/energy relation by considering a massless
supergravity probe in the KT background. For a massless scalar field the
equation of motion is
\begin{equation} \label{sugraeom}                   
\nabla^2 \phi = 
\frac{1}{L^2 \sqrt{\ln (r/r_s)}} 
\left[ \frac{L^4 \ln (r/r_s)}{r^2} 
\partial_i \partial^i +\frac{1}{r^3} 
\frac{\partial}{\partial r} r^5 \frac{\partial}{\partial r} \right] \phi =0 \ .
\end{equation}
The second term is invariant under a rescaling of the radius.  Thus  a solution 
of (\ref{sugraeom}) which is wavelike on a radial
slice of $AdS$ depends on the radius through the quantity $\frac{L^4 \ln
(r/r_s)k^2}{r^2}$.  For this prescription
\begin{equation} \label{sugracutoff}                  
   \Lambda \sim \frac{r_c}{L^2\sqrt{\ln (r_c/r_s)}} \ .
\end{equation}
We can obtain the same result by another physical argument which
seems quite different on its surface.  Let us consider instead this
theory at finite temperature, as was studied in \cite{finitetemp}.  At
sufficiently high temperature, the system develops a horizon, and the
Hawking temperature is related to the horizon radius $r_H$ by
\begin{equation}
   T_H \sim \frac{r_H}{L^2 \sqrt{\ln(r_H/r_s)}} \ , 
\end{equation}
in the limit of high temperature.  In the theory with a large
radius cutoff, the maximum temperature is simply given by setting
$r_H=r_c$.  Then if we identify the UV cutoff $\Lambda$ as the maximum
Hawking temperature, we recover the result (\ref{sugracutoff}).  The
agreement between these two methods for relating the RG scale to the
cutoff radius is a sign that holography is at work.

Let us note that
the relations we find between $\mu$ and $\Lambda$ are {\it exactly}
of the form one finds in an
asymptotically free gauge theory. 
To make contact with the standard dimensional
transmutation formula, we have to identify
 $\tau_c/3 = 8\pi^2/( \beta_0 g_0^2)$.  
If the beta function is
\[
\frac{dg}{d \log(\Lambda/\mu) } = 
-\beta_0 \frac{g^3}{16\pi^2}  - \beta_1 \frac{g^5}{128 \pi^4} - \ldots
\ ,
\]
then we find
\be
\mu \sim \Lambda e^{-8 \pi^2/(\beta_0 g_0^2)} 
(8 \pi^2/\beta_0 g_0^2)^{\beta_1/\beta_0^2} \ .
\ee
To  make contact with the SUGRA result (\ref{gravdim})
we have to identify
 $\tau_c/3 = 8\pi^2/( \beta_0 g_0^2)$.

We can reexpress $\tau_c$ in terms of the NS-NS flux
\[
\int_{\tau < \tau_c} \int_{S^2} H_3 = 4 \pi^2 \alpha' K \ .
\]
One quickly finds 
$\tau_c \approx 2 \pi K/(g_s M)$.
In order to achieve the continuum limit, we have to take 
$\tau_c \rightarrow \infty$, $\Lambda\rightarrow\infty$
while keeping the physical scale $\mu$ fixed.
The exponential factor $e^{-\tau_c/3} = e^{-2\pi K/(3 g_s M)}$,
which may give rise to a large hierarchy of scales in 
compactified F-theory, was derived in \cite{GKP}.
It was observed that the type IIB
supergravity picture of gluino condensation is
reminiscent of the gluino condensation in the hidden sector
of the heterotic string \cite{DSRW}. 
 Here we note that a more precise
SUGRA analysis may produce a power-law prefactor,  which 
is analogous to the prefactor due to the 2-loop
$\beta$-function in an asymptotically free
gauge theory.  
With the stretched-string relation, we 
find $\beta_1/\beta_0^2=0$; the 
redshift relation gives $\beta_1/\beta_0^2=1/4$; 
and the holographic relation gives $\beta_1/\beta_0^2=1/2$.

{}For pure ${\cal N}=1$ supersymmetric 
gauge theory with gauge group some simple Lie group $G$, 
$\beta_0 = 3 C_2(G)$ and $\beta_1 = 3 C_2(G)^2$
(see for example \cite{NSVZ}).  The quantity $C_2(G)$ is the quadratic 
Casimir.
Unfortunately, none of our distance/energy relations give the required value of 1/3.  Perhaps adding the right kind of matter will fix the ratio to the 
correct value. It is also possible that a different identification
between the SUGRA and field theory couplings may fix the prefactor.
In any case,
it would be very interesting to find out if the analogy with the gluino
condensation in the hidden sector of the heterotic string \cite{GKP}
is in fact an exact duality. 

\appendix

\section{Volume of the Two and Three Cycles}

The manifold $T^{1,1}$ can be identified as the intersection of the conifold
$\sum_{i=1}^4 z_i^2 = 0$ 
and the sphere
$\sum_{i=1}^4 |z_i|^2 = 1\ , $
where $z_i \in \C{}$.  Dividing up $z_i = x_i + i y_i$ into real and imaginary
parts, we see that $T^{1,1}$ can be thought of as the set of points 
satisfying $\sum x_i^2 = 1/2$ and $\sum y_i^2 = 1/2$ along with the 
constraint ${\bf x}\cdot{\bf y}=0$.  If we use this constraint to 
eliminate one of the $x$, we can see, at least in a heuristic way, 
that the manifold $T^{1,1}$ can be thought of as a $\S^2$ defined 
by the remaining $x_i$ fibered over an $\S^3$ base defined by the $y_i$.

We now use this observation to parametrize the two cycle $C_2$.  An explicit
parametrization of the whole $T^{1,1}$ is known in terms of the angles 
$0 \leq \psi < 4 \pi$, $0 \leq \theta_i \leq \pi$, and 
$0 \leq \phi_i < 2 \pi$ where $i = 1,2$.  Indeed
\[
z_1 = \frac{e^{i\psi/2}}{\sqrt{2}} \left(
\cos \left(\frac{\theta_1+\theta_2}{2} \right)
\cos \left(\frac{\phi_1 + \phi_2}{2} \right) + 
i \cos \left(\frac{\theta_1-\theta_2}{2} \right)
\sin \left(\frac{\phi_1 + \phi_2}{2} \right)
\right)
\]
\[
z_2 = \frac{e^{i\psi/2}}{\sqrt{2}} \left(
-\cos \left(\frac{\theta_1+\theta_2}{2} \right)
\sin \left(\frac{\phi_1 + \phi_2}{2} \right) + 
i \cos \left(\frac{\theta_1-\theta_2}{2} \right)
\cos \left(\frac{\phi_1 + \phi_2}{2} \right)
\right)
\]
\[
z_3 = \frac{e^{i\psi/2}}{\sqrt{2}} \left(
-\sin \left(\frac{\theta_1+\theta_2}{2} \right)
\cos \left(\frac{\phi_1 - \phi_2}{2} \right) + 
i \sin \left(\frac{\theta_1-\theta_2}{2} \right)
\sin \left(\frac{\phi_1 - \phi_2}{2} \right)
\right)
\]
\[
z_4 = \frac{e^{i\psi/2}}{\sqrt{2}} \left(
-\sin \left(\frac{\theta_1+\theta_2}{2} \right)
\sin \left(\frac{\phi_1 - \phi_2}{2} \right) - 
i \sin \left(\frac{\theta_1-\theta_2}{2} \right)
\cos \left(\frac{\phi_1 - \phi_2}{2} \right)
\right)
\]
To describe the fiber, we would like to stay on one point on the base $\S^3$.  Thus, we want to keep the imaginary part of the $z_i$ constant while still keeping two degrees of freedom available to trace out the $\S^2$ fiber.  For convenience, we begin by
 choosing $\psi=0$.  From the parametrization, we can trace out the $\S^2$ by setting $\theta_1 = \theta_2$ and $\phi_1 = -\phi_2$.
Integrating using these coordinates, $\int_{C_2} \omega_2 = 4 \pi$. 

Next we consider the three cycle $C_3$.  First recall that
\be
g^5 \wedge g^3 \wedge g^4 = \omega_3 - \frac12 d(g^1 \wedge g^3 + g^2 \wedge g^4) \ .
\ee
Moreover, $C_3$ has no boundary so 
\be
\int_{C_3} \omega_3 = \int_{C_3} g^5 \wedge g^3 \wedge g^4 \ .
\ee
We recall from \cite{MT} that 
\be
ds^2 = \frac{1}{2} (g^5)^2 + (g^3)^2 + (g^4)^2 
\ee
is the standard metric on a $\S^3$ with radius $\sqrt{2}$.  Moreover
${\rm Vol}(\S^3) = 2 \pi^2 r^3$.  It follows then that 
$\int_{C_3} \omega_3 = 8 \pi^2$.

\section{Complex notation for the forms}

Our strategy is to guess differential forms, written in terms
of the $z_i$, with the appropriate 
symmetries and properties.  To refine further and check the guess,
we use computer assisted algebra to rewrite the differential forms 
in terms of the angular 
coordinates on the conifold.  We can then compare the guess
with the differential forms given in previous sections in terms
of the $g^i$. 

{}First, we need to 
review the construction of the angular coordinates on the
deformed conifold.
We define
\be
W = \left(
\begin{array}{cc}
z_3 + iz_4 & z_1 - iz_2 \\
z_1 + iz_2 & -z_3 + iz_4 
\end{array}
\right) \ .
\ee
The defining relation of the 
deformed conifold (\ref{dconifold}) becomes
$\det W = - \varepsilon^2$.
We introduce the angular coordinates with two $SU(2)$ $j=1,2$ matrices
\be
L_j = \left(
\begin{array}{cc}
\cos \frac{\theta_j}{2} \, e^{i(\psi_j + \phi_j)/2}& 
-\sin \frac{\theta_j}{2} \, e^{-i(\psi_j - \phi_j)/2}\\
\sin \frac{\theta_j}{2} \, e^{i(\psi_j - \phi_j)/2}&  
 \cos \frac{\theta_j}{2} \, e^{-i(\psi_j + \phi_j)/2}
\end{array}
\right) \ .
\ee
The idea is then to take some representative point $p \in C$ corresponding to
\be
W_0 = \left(
\begin{array}{cc}
0 & \varepsilon e^{\tau/2} \\
\varepsilon e^{-\tau/2} & 0
\end{array}
\right) \ .
\ee
We can generate all of $C$ by acting on $W_0$ with $L_1$ and $L_2$
\be
W = L_1 \cdot W_0 \cdot L_2^\dagger \ .
\ee
As the coordinates $\psi_1$ and $\psi_2$ only appear in $W$ as 
$\psi_1 + \psi_2$, we may define a new variable $\psi = \psi_1 + \psi_2$.
It is natural to define a radial coordinate 
\be
\rho^2 \equiv \sum_{i=1}^4 z_i \bar{z}_i = \frac{1}{2} \Tr (W \cdot W^\dagger)
\ .
\ee
With this definition, one straightforwardly obtains
$\rho^2 = \varepsilon^2 \cosh \tau$.
The singular case, $\varepsilon = 0$ is recovered by taking the large $\tau$ limit.  
Equivalently, we may start with a slightly different 
\be
W_0^{sing} = 
\left(
\begin{array}{cc}
0 & \sqrt{2} \rho \\
0 & 0
\end{array}
\right) \ .
\ee
To summarize, then, 
the angular coordinates on the conifold are $0 \leq \psi < 4\pi$, 
$0 \leq \theta_j < \pi$, $0 \leq \phi_j < 2 \pi$ and a radius $\rho$.  In
the case where $\varepsilon \neq 0$, we can substitute $\tau$ for $\rho$.
In the case $\varepsilon=0$, $\rho$ is typically expressed as 
$r^3 \sim \rho^2$
in order to make the conical nature of the metric evident (\ref{conimetric}).

In principle, we have an explicit coordinate transformation between the 
angular variables and the complex coordinates $z_i$.  The goal of this appendix,
to rewrite the important supergravity quantities in terms of the $z_i$, should
be a straightforward task.
Given any quantity written in terms of the angles, we should be able to 
write down the same quantity in terms
of the $z_i$.  In practice, this variable change is difficult for two reasons.
First, and especially in the case $\varepsilon \neq 0$, the variable change 
is quite complicated and nearly impossible to do without some computer assisted
algebra.  Second, there are more $z_i$ than one needs.  By choosing three out of
the four $z_i$, one explicitly breaks the SO(4) symmetry.  The formulae involving 
only three $z_i$ are typically messy and uninformative.  Moreover, it 
is usually not completely obvious how to reintroduce the fourth $z_i$ in a way
that symmetrizes the quantity of interest.

\subsection{Forms on the Singular Conifold}

We begin with the easier case, the singular conifold (\ref{coni}). 
We would like to express the forms important to the KT solution
\cite{KT} and discussed in section 2 in terms of the $z_i$.
It is a fact that $g^5$, $\omega_2$, and $\omega_3$ all transform
as singlets under the $SO(4)$ action.
Another important symmetry that holds
when $\varepsilon=0$ is the scaling $z_i \to \lambda z_i$ where 
$\lambda \in \C{*}$.  The real part of this scaling, i.e. 
$\lambda \in \R{+}$, corresponds to scaling the radius $\rho$, while the 
complex U(1) part, i.e. $\lambda = e^{i\alpha}$, 
corresponds to shifting the angle
$\psi$.
Cursory inspection of the vielbeins and the defining relations
for $\omega_2$ and $\omega_3$ (\ref{fbasis}, \ref{omegato}, \ref{omegathr})
show that the forms $g^5$, $\omega_2$, and
$\omega_3$ are invariant under the full scaling.

Begin with $g^5$.  Using the $z_i$, there are two ways to write down
singlet one forms which obey the scaling symmetries: 
$z_i \, d\bar{z}_i / \rho^2$ or $\bar{z}_i \, dz_i / \rho^2$ where
summation on the indices is implied.\footnote{
Note that $z_i \, dz_i$ and its complex conjugate 
vanish by the defining relation
on the conifold (\ref{dconifold}).}
Any singlet one form must be some 
linear combination of these two, and all that need be done is
fix the constants.  Indeed
\be
\frac{d\rho}{\rho} + \frac{i}{2} g^5 = \frac{1}{\rho^2} \bar{z}_i \, dz_i 
\label{scof}
\ee
and 
\be
\frac{d\rho}{\rho} - \frac{i}{2} g^5 = \frac{1}{\rho^2} z_i \, d\bar{z}_i \ .
\ee

Next, we consider the two form $\omega_2$.  There are several ways of writing 
SO(4) invariant two forms.  Indeed
\[
\eta_1 = \epsilon_{ijkl} \, z_i \bar{z}_j \, dz_k \wedge d\bar{z}_l 
\ , \; \; \; \;
\eta_2 = \epsilon_{ijkl} \, z_i \bar{z}_j \, dz_k \wedge dz_l \ ,
\]
\[
\eta_3 = \epsilon_{ijkl} \, z_i \bar{z}_j \, d\bar{z}_k \wedge d\bar{z}_l \ , \;\;\;\;
\eta_4 = (z_i \, d\bar{z}_i) \wedge (\bar{z}_j \, dz_j) \ , \;\;\;\;
\]
\be
\eta_5 = (dz_i \wedge d\bar{z}_i) \ .
\label{etas}
\ee
We can eliminate $\eta_2$ and $\eta_3$ immediately because they 
explicitly break the $U(1)$ symmetry $z_i \to e^{i\alpha} z_i$.
The situation 
is even simpler.  The form $\omega_2$ transforms with a minus sign under the
spatial inversion $z_1 \to -z_1$, keeping all the other $z_i$ fixed, while the
forms $\eta_4$ and $\eta_5$ are invariant under the full $O(4)$ symmetry.  
Our constraints leave only $\eta_1$ as a candidate for $\omega_2$:  
\be
\omega_2 = \frac{i}{\rho^4} \eta_1 \ .
\label{sctf}
\ee
As mentioned in \cite{KWnew}, 
this form is closed, as we may check explicitly:
\be
\partial\omega_2 = -\frac{i}{\rho^6} \left(
2 \chi_1
+ \rho^2 \, \chi_4
\right) \ ,
\label{partot}
\ee
where
\be
\chi_1 = (\epsilon_{ijkl} \, z_i \bar{z}_j \, 
dz_k \wedge d\bar{z}_l) \wedge (\bar{z}_m \, dz_m) \ , \; \; \; \;
\chi_4 = \epsilon_{ijkl} \, \bar{z}_i \, dz_j \wedge dz_k \wedge d\bar{z}_l \ ,
\ee
are new $(2,1)$ $SO(4)$ invariant forms, labeled to conform with
the notation used in the next section.
With computer assisted algebra, using (\ref{coni}), 
it is easy to see that the expression on the
right side of (\ref{partot}) 
vanishes.  We also provide a symmetry argument for the 
vanishing.  Because $\chi_1$ and $\chi_4$
are $SO(4)$ and scale invariant, we are free to check the vanishing
for a specific point $p$ on 
the singular conifold, $z_1 = 1$, $z_2=i$, 
$z_3=0$, and $z_4=0$, and
then invoke the $SO(4)$ and scaling 
symmetry to prove the vanishing for all points.

{}Finally, we turn to $\omega_3$, and in fact we already know the answer because
$\omega_3 = g^5 \wedge \omega_2$.  The most important form in the KT solution
is not $F_3$ or $H_3$ (and hence $\omega_3$ or $\omega_2$) independently 
but their combination $G_3 = F_3 - i H_3 / g_s$, which needs to be a 
$(2,1)$ form in order to preserve supersymmetry.  We may check that
\be
G_3 = \frac{\alpha'}{2} M \left( g^5 - \frac{2 i d\rho}{\rho}\right)
\wedge \omega_2  =
\frac{\alpha'}{\rho^6} M 
\left( \epsilon_{ijkl} \, z_i \bar{z}_j \, dz_k \wedge d\bar{z}_l \right) \wedge
\left( \bar{z}_m \, dz_m \right)  
\label{scthrf}
\ee
which is explicitly a $(2,1)$ form, as required.  Reassuringly, 
changing the sign of $d\rho / \rho$
in the expression above produces instead a $(1,2)$ form.

\subsection{Forms on the Deformed Conifold}

The deformed conifold is more difficult, not only because the coordinate 
transformation is more complicated but also because the nonzero $\varepsilon$
explicitly breaks the $U(1)$ and scale invariance.

Before tackling the (2,1) form $G_3 = F_3 - i H_3 /g_s$, 
let us warm up with some
simpler one and two forms.  First, we check what happens to 
$z_i \, d\bar{z}_i$ and
$\bar{z}_i \, dz_i$ when $\varepsilon$ is turned on:
\be
\bar{z}_i \, dz_i = 
\varepsilon^2 \frac{\sinh \tau}{2} \left( d\tau + ig^5 \right) \ .
\ee
Comfortingly, in the large $\tau$ limit, we recover the singular
conifold result (\ref{scof}).  The result for $z_i \, d\bar{z}_i$ is, 
not surprisingly, the complex conjugate.

Next we consider what happens to $\omega_2$.  Remember that we have broken
the $U(1)$ symmetry and as a result, the forms $\eta_2$ and $\eta_3$ 
(\ref{etas}) may contribute.  Indeed, they do as
\be
\omega_2 = \frac{i \cosh \tau}{\varepsilon^4 \sinh^3 \tau} \epsilon_{ijkl} \, z_i \bar{z}_j
\left(
dz_k \wedge d\bar{z}_l - \frac{1}{2\cosh \tau} \left(
dz_k \wedge dz_l + d\bar{z}_k \wedge d\bar{z}_l
\right)
\right) \ .
\ee
In the large $\tau$ limit, the reader may check that we recover 
the result for the singular conifold (\ref{sctf}).

Although $\omega_2$ becomes a combination of different U(1) breaking
differential forms, the NS-NS potential $B_2$ is actually more closely
related to the old $\omega_2$ of the singular conifold.  Indeed
\be
B_2 = \frac{i g_s M \alpha'}{2 \varepsilon^4} 
\frac{\tau \coth \tau - 1}{\sinh^2 \tau}
\eta_1 \ .
\ee
Again, in the large $\tau$ limit, we recover $B_2$ on the singular conifold.

Now we are ready to attack the harmonic $(2,1)$ form $G_3$.  We begin by writing
down all of the $SO(4)$ invariant $(2,1)$ forms, of which there are five,
\[
\chi_1 = (\epsilon_{ijkl} \, z_i \bar{z}_j \, dz_k \wedge d\bar{z}_l)
\wedge (\bar{z}_m \, dz_m)
\ ,
\]
\[ 
\chi_2 = (\epsilon_{ijkl} \, z_i \bar{z}_j \, dz_k \wedge dz_l)
\wedge (z_m \, d\bar{z}_m) \ ,
\]
\[
\chi_3 = (\epsilon_{ijkl} \, z_i \, dz_j \wedge dz_k \wedge d\bar{z}_l)
\ , 
\]
\[
\chi_4 = (\epsilon_{ijkl} \, \bar{z}_i \, dz_j \wedge dz_k \wedge d\bar{z}_l) 
\ , 
\]
\be
\chi_5 = (dz_i \wedge d\bar{z}_i) \wedge (\bar{z}_j \, dz_j) \ .
\ee
{}Fortunately we can eliminate $\chi_5$ because
$\partial \left( h(\rho^2) \eta_4 \right) = h(\rho^2) \chi_5$.  
Based on experience with the singular conifold,
and in particular the demonstration that $\partial \omega_2=0$, one
may wonder if the remaining $\chi_i$ are linearly independent on the deformed
conifold.  They are not.  The equation
\[
\chi \equiv \alpha \chi_1 + \beta \chi_2 + \gamma \chi_3 + \delta \chi_4 =0 
\]
is easy to satisfy.  We choose a convenient point $p \in C$, for example
the point corresponding to the matrix $W_0$.  If $\chi$ vanishes
at $p$, it vanishes on all of $C$ by $SO(4)$ invariance.  The two conditions
that must be met for $\chi$ to vanish are
\[
\alpha \cosh \tau + 2 \beta - 2 \delta / \varepsilon^2 = 0 \ ,
\]
\be
\alpha + 2\beta \cosh \tau + 2 \gamma / \varepsilon^2 = 0 \ .
\label{lindep}
\ee

We choose the ansatz for the $(2,1)$ form 
\be
G_3 = \alpha \chi_1 + \beta \chi_2 + \gamma \chi_3 + \delta \chi_4 \ .
\ee
An intensive computer assisted computation reveals 
\[
\alpha \cosh \tau + 2 \beta - 2 \delta / \varepsilon^2 =
\frac{M\alpha'}{2 \varepsilon^6 \sinh^5 \tau}
\left[ \sinh \tau \left( \cosh (2\tau) + 5 \right) - 6 \tau \cosh \tau
\right] \ ,
\]
\be
\alpha + 2\beta \cosh \tau + 2 \gamma / \varepsilon^2 = 
-\frac{M \alpha' }{\varepsilon^6 \sinh^5 \tau}
\left[
\tau \left( \cosh (2\tau) + 2 \right) - 3 \sinh \tau \cosh \tau
\right] \ .
\label{toformeq}
\ee
Because of the linear dependence of the $\chi_i$, (\ref{lindep}), 
we are free to choose any two
of the four parameters $\alpha$, $\beta$, $\gamma$, and $\delta$ freely.
Said another way, we can express $G_3$ as the sum of any two $\chi_i$, 
$i=1, \ldots, 4$.

Let us choose $\gamma=0$ and $\delta=0$.  In this case,
\[
\alpha = \frac{M \alpha'}{2\varepsilon^6} 
\frac{\sinh (2\tau)-2\tau}{\sinh^5 \tau}
\]
and
\[
\beta = \frac{M \alpha'}{2\varepsilon^6} 
\frac{2 (1 -\tau \coth \tau)}{\sinh^4 \tau} \ .
\]
In the large $\tau$ limit, $\beta$ becomes vanishingly small 
compared to $\alpha$.  If it did not, the $U(1)$ symmetry on the 
singular conifold would not be preserved!  Moreover, 
$\alpha \to M \alpha' / \rho^6$, in agreement with (\ref{scthrf}).

\section*{Acknowledgements}
I.R.K. is grateful to S. Gubser, N. Nekrasov, M. Strassler,
A. Tseytlin and E. Witten for collaboration on parts of the material 
reviewed in these notes and for useful input. 
We also thank Sergey Frolov and John Pearson for 
useful discussions.
This work was supported in part by the NSF grant PHY-9802484.


\end{document}